\documentclass[fleqn,usenatbib]{mnras}

\usepackage[T1]{fontenc}
\usepackage{ae,aecompl}

\usepackage{graphicx}	
\usepackage{amsmath}	
\usepackage{amssymb}	
\usepackage{epstopdf}

\usepackage{newtxtext,newtxmath}

\usepackage{color}
\definecolor{vero}{rgb}{0.6, 0.4, 0.8}

\definecolor{newalex}{rgb}{0.0, 0.6, 0.0}

\title[MOBSTER - IV. New magnetic B star]{MOBSTER - IV. Detection of a new magnetic B-type star from follow-up spectropolarimetric observations of photometrically selected candidates\thanks{Based on observations collected at the Canada-France-Hawaii Telescope (CFHT).}}

\author[A. David-Uraz et al.]{A. David-Uraz,$^{1,2,3}$\thanks{E-mail: alexandre.daviduraz@howard.edu}
M.~E. Shultz,$^{1}$
V. Petit,$^{1}$
D.~M. Bowman,$^{4}$ C. Erba,$^{1}$\newauthor R.~A. Fine,$^{1}$ C. Neiner,$^{5}$ H. Pablo,$^{6}$ J. Sikora,$^{7}$ A. ud-Doula$^{8}$, and G.~A. Wade$^{9}$
\\
$^{1}$Department of Physics and Astronomy, and Bartol Research Institute, University of Delaware, Newark, DE 19716, USA\\
$^{2}$Department of Physics and Astronomy, Howard University, Washington, DC 20059, USA\\
$^{3}$Center for Research and Exploration in Space Science and Technology, and X-ray Astrophysics Laboratory, NASA/GSFC, Greenbelt, MD 20771, USA\\
$^{4}$
{Institute of Astronomy}, KU Leuven, Celestijnenlaan 200D, B-3001 Leuven, Belgium\\
$^{5}$LESIA, Paris Observatory, PSL University, CNRS, Sorbonne University, Universit\'{e} de Paris, 5 place Jules Janssen,\\ 92195 Meudon, France\\
$^{6}$American Association of Variable Star Observers, 49 Bay State Rd., Cambridge, MA, 02138, USA\\
$^{7}$Department of Physics and Astronomy, Bishop's University, Sherbrooke, Qu{\'e}bec, Canada, J1M 1Z7\\
$^{8}$Penn State Scranton, 120 Ridge View Drive, Dunmore, PA 18512, USA\\
$^{9}$Department of Physics and Space Physics, Royal Military College of Canada, PO Box 17000 Kingston, ON K7K 7B4, Canada
}

\date{Accepted 2021 March 23. Received 2021 March 19; in original form 2020 April 21}

\pubyear{2020}

\begin{document}
\label{firstpage}
\pagerange{\pageref{firstpage}--\pageref{lastpage}}
\maketitle

\begin{abstract}
In this paper, we present results from the spectropolarimetric follow-up of photometrically selected candidate magnetic B stars from the MOBSTER project. Out of four observed targets, one (HD 38170) is found to host a detectable surface magnetic field, with a maximum longitudinal field measurement of 105$\pm$14 G. This star is chemically peculiar and classified as an $\alpha^2$ CVn variable. Its detection validates the use of \textit{TESS} to perform a photometric selection of magnetic candidates. Furthermore, upper limits on the strength of a putative dipolar magnetic field are derived for the remaining three stars, and we report the discovery of a previously unknown spectroscopic binary system, HD 25709. Finally, we use our non-detections as case studies to further inform the criteria to be used for the selection of a larger sample of stars to be followed up using high-resolution spectropolarimetry.
\end{abstract}

\begin{keywords}
stars: early-type -- stars: magnetic field -- stars: rotation -- techniques: photometric -- techniques: spectroscopic -- techniques: polarimetric
\end{keywords}



\section{Introduction}

It has been well established by large spectropolarimetric surveys that there exists a distinct population of magnetic OBA stars, with an incidence of roughly 10 
{per cent} \citep[e.g.][]{2009ARA&A..47..333D, 2016MNRAS.456....2W, 2017MNRAS.465.2432G, 2017A&A...599A..66S, 2019MNRAS.483.3127S}. 
{A key feature of this population is} a lack of correlation between magnetic and stellar properties 
\citep{2019MNRAS.490..274S}, and there is still much debate surrounding the formation mechanism of these magnetic fields \citep[e.g.][]{2016MNRAS.457.2355S, 2019A&A...622A..72V}. 
{The evolution of the magnetic field} is also poorly constrained, although there are preliminary indications that the usual assumption of flux conservation might not hold at higher masses {\citep{2007A&A...470..685L,2008A&A...481..465L,2016A&A...592A..84F,2019MNRAS.490..274S}}. 

The main hindrance to {answering} these fundamental questions resides in small number statistics, especially for {the} more massive {magnetic} stars: to date, there are only 11 known magnetic O-type stars \citep{2013MNRAS.429..398P, 2016A&A...592A..84F} and less than 100 known magnetic {early} B-type stars \citep{2018MNRAS.475.5144S}. Given the 
{large} cost of spectropolarimetry, {we propose that} the potential of 
{blind} surveys to significantly increase this sample has {effectively} reached its limit{. 
Targeted} efforts relying on indirect magnetic diagnostics to build up candidate lists yield much higher detection rates \citep{2018MNRAS.478.2777B}, although for fainter stars, the confirmation of their magnetic status might require the development of next-generation spectropolarimeters equipped on 10m-class telescopes. Such diagnostics include H$\alpha$ or Brackett line emission \citep[e.g.][]{1974ApJ...191L..95W, 2014ApJ...784L..30E}, rotationally modulated photometry \citep[e.g.][]{2020MNRAS.492.1199M, 2020A&A...635A.163B}, chemical peculiarities and rotational modulation due to chemical spots \citep{1958ApJ...128..228B} and the presence of specific spectral features such as the peculiar C\textsc{iii}/N\textsc{iii} {emission} complex in the {optical spectra of} Of?p spectral class {stars} \citep{1972AJ.....77..312W, 2017MNRAS.465.2432G}. Additionally, 
high-precision photometric missions such as the Transiting Exoplanet Survey Satellite (\textit{TESS}{;} \citealt{2015JATIS...1a4003R}) 
play an instrumental role in identifying such magnetic candidates, through the detection of periodic signals associated with rotation. The MOBSTER (Magnetic OB[A] Stars with \textit{TESS}: probing their Evolutionary and Rotational properties) Collaboration\footnote{This collaboration seeks to characterize the photometric variability of known magnetic stars (e.g. \citealt{2019MNRAS.487..304D}) and conversely to identify magnetic candidates from photometry, with the eventual goal of determining the rotational (e.g. \citealt{2019MNRAS.487.4695S}) and evolutionary (e.g. \citealt{2019MNRAS.490.4154S}) properties of the subpopulation of OBA stars with detectable magnetic fields on their surfaces.} was formed to leverage \textit{TESS} for 
{this} specific purpose.

This paper presents the first new magnetic detection achieved by the MOBSTER Collaboration and establishes the bases for its ongoing efforts to perform 
targeted spectropolarimetric 
{surveys} of massive and intermediate-mass magnetic candidates. In Section~\ref{sec:phot}, we describe the photometric observations that were used and the analysis that was performed to select candidates. Section~\ref{sec:specpol} describes the follow-up spectropolarimetry of four of these candidates and details the {results} of our magnetometric analysis. Finally, in Section~\ref{sec:conclusions}, we discuss our results and draw conclusions about 
{the} directions our work 
{will} take moving forward.

\section{Photometry}\label{sec:phot}


\subsection{Observations} 

Photometric observations were obtained by \textit{TESS}. In the context of this mission, the sky is divided {into} ``sectors" that are observed for 27.4 d. Most objects lie within only one sector, although there is some overlap between the sectors (especially toward the ecliptic poles) 
{leading} {to} some objects 
{being} observed for a longer temporal baseline. 

The \textit{TESS} bandpass is broad, covering wavelengths between about 600 and 1000 nm. While about half a billion point sources are included in the \textit{TESS} Input Catalogue (TIC; \citealt{2018AJ....156..102S, 2019AJ....158..138S}) and thus appear in the {30-min} cadence full-frame images ({FFIs}), the Candidate Target List (CTL) contains a subset of {priority} objects to be observed in {2-min} cadence. The \textit{TESS} Science Processing Operations Center (SPOC; \citealt{2016SPIE.9913E..3EJ}) releases reduced light curves based on the {2-min} cadence data 
{at} the Mikulski Archive for Space Telescopes (MAST)\footnote{\url{https://archive.stsci.edu/missions-and-data/transiting-exoplanet-survey-satellite-tess}}. While this reduction is optimized for 
{exoplanet detection} and certain {stellar} signals might {therefore} be suppressed (especially longer{-duration} signals in hot stars, with periods greater than several days, e.g. \citealt{2020pase.conf..226B}), it remains a useful resource for an efficient first-pass effort at identifying candidate {magnetic 
stars}. However, its usefulness might be biased towards shorter periods; therefore, reprocessing of 30-min cadence data will be necessary, especially for earlier-type magnetic stars, which tend to have longer rotation periods due to magnetic spindown, \citealt{2009MNRAS.392.1022U}).

\subsection{Target selection}

\begin{figure*}
	\includegraphics[width=0.49\linewidth]{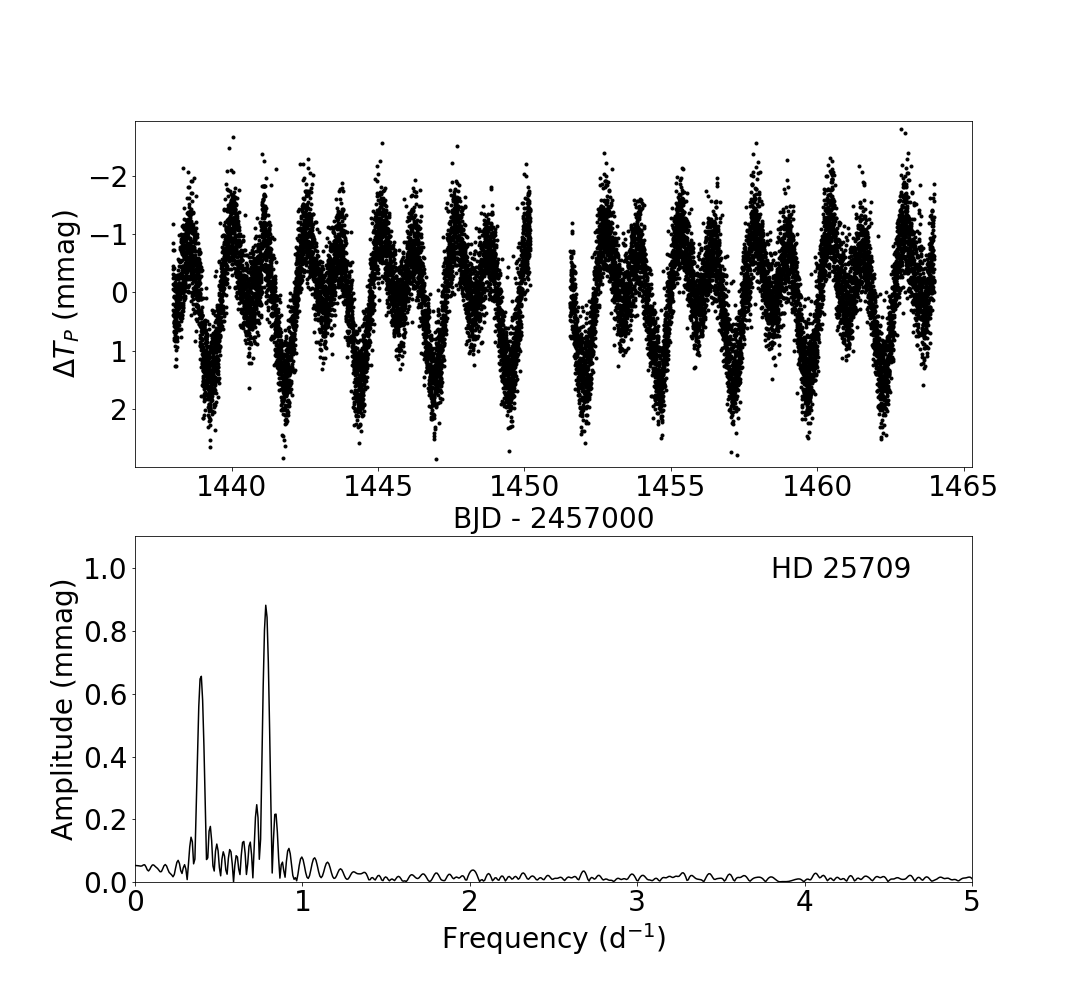}
	\includegraphics[width=0.49\linewidth]{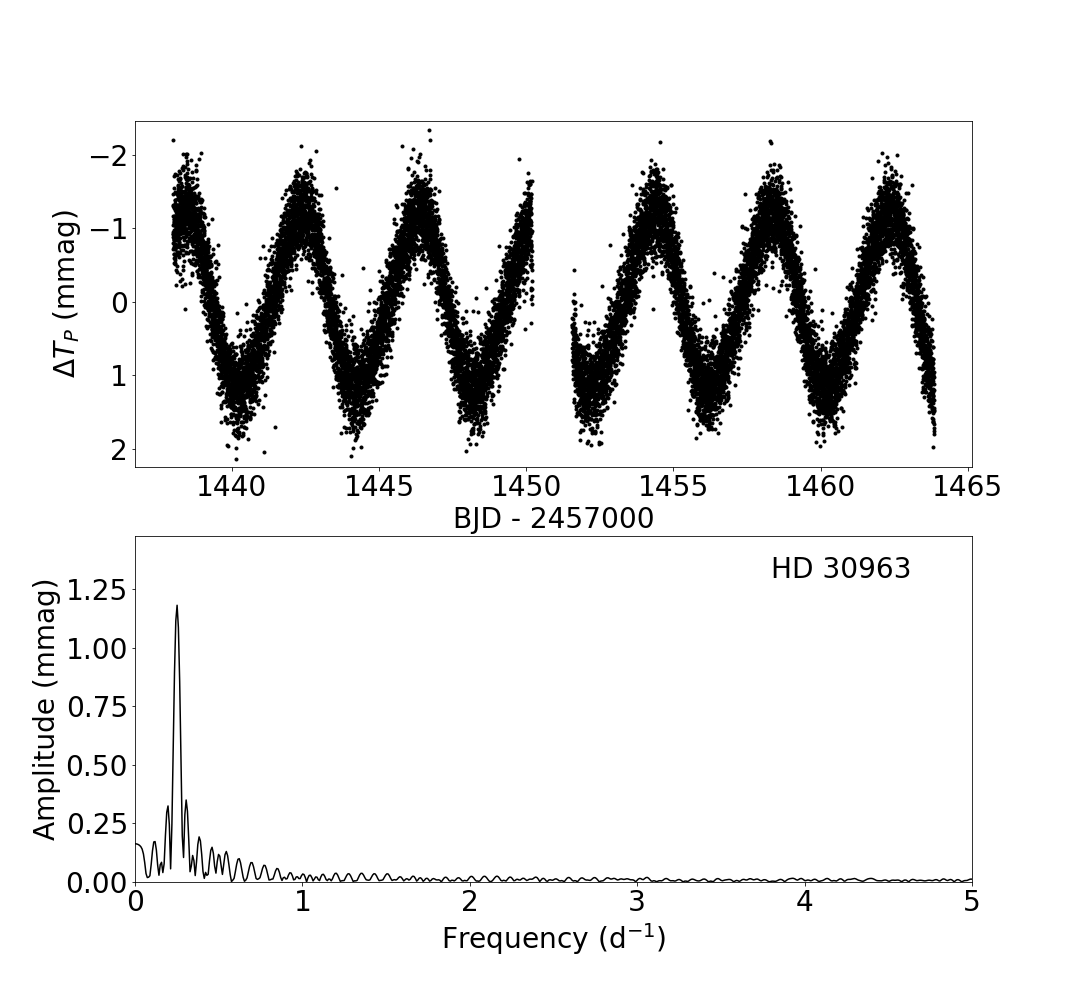}
	\includegraphics[width=0.49\linewidth]{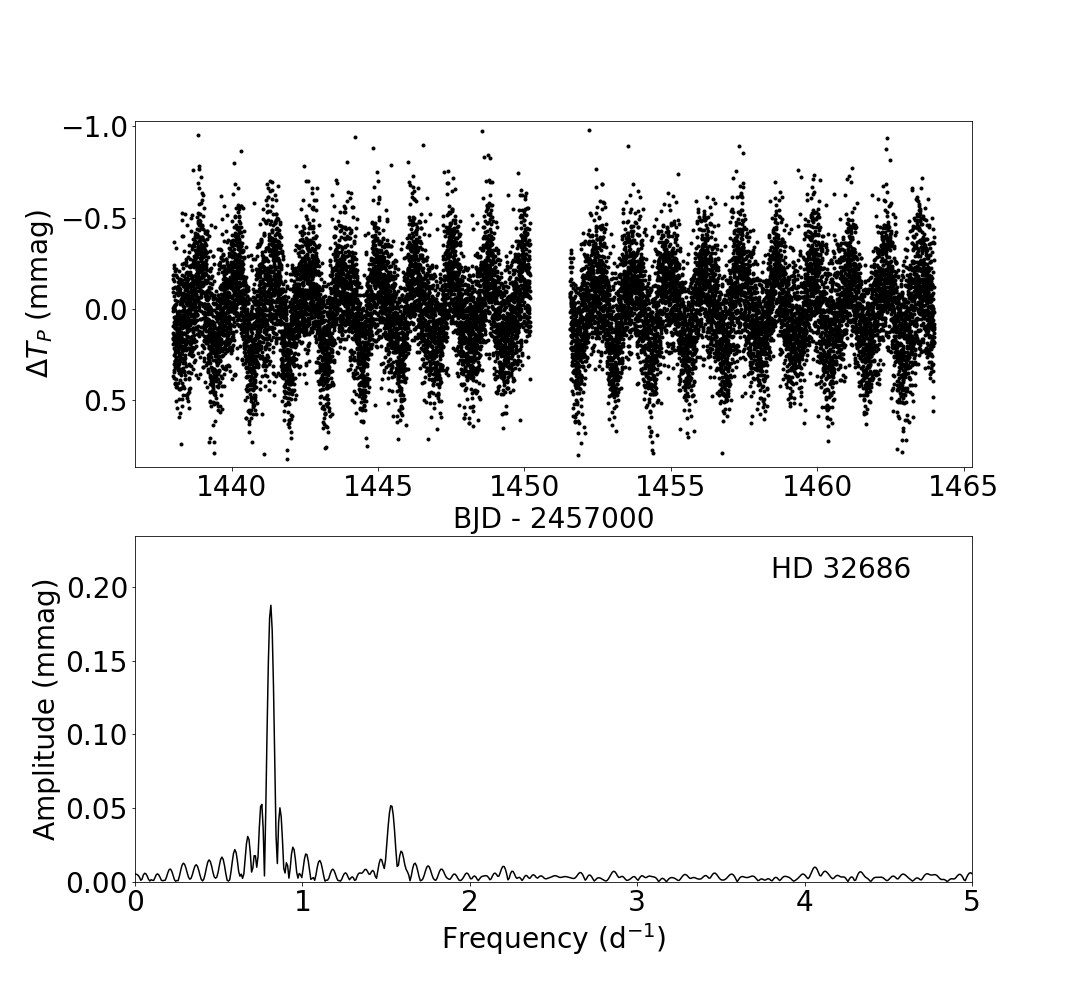}
	\includegraphics[width=0.49\linewidth]{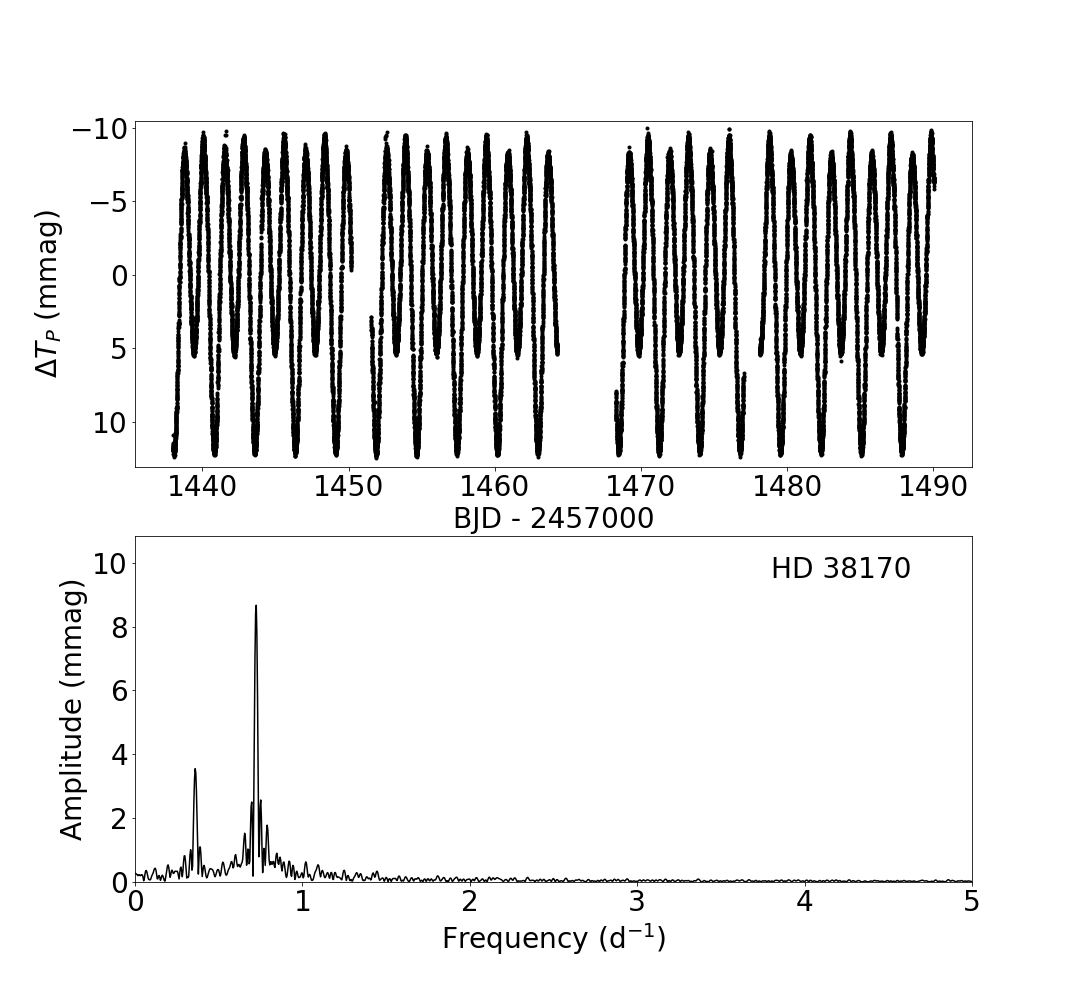}
    \caption{Detrended \textit{TESS} light curves (differential \textit{TESS} instrumental magnitude, on the y-axis, with the zero-point corresponding to the average brightness, vs. date on the x-axis) and associated Lomb-Scargle periodograms for all four targets selected for spectropolarimetric follow-up and observed by CFHT. Note that the latter are given in amplitude rather than power (with the amplitude $A$ given by $A = 2\sqrt{P/N}$, where $P$ is the unnormalized power output by the \texttt{autopower} function of the \texttt{astropy.stats.LombScargle} class in \textsc{python} and $N$ is the number of points in the light curve) to enable easier comparison and signal identification with the light curves above.}
    \label{fig:lc}
\end{figure*}

We retrieved 
{2-min} cadence light curves {from MAST} for objects identified as B-type stars 
{in} the SIMBAD database that were observed by \textit{TESS} in Sectors 1 to 6. We then performed a Lomb-Scargle {frequency} analysis of these light curves \citep{1976Ap&SS..39..447L, 1982ApJ...263..835S} and identified potential rotational modulation signals in a subset of them based on the criteria laid out by \citet{2013MNRAS.431.2240B} and adopted by \citet{2019MNRAS.487.4695S}
{: specifically,} the detection of a significant peak within a range of frequency consistent with rotation (less than about 4 d$^{-1}$, based on the range of critical rotation rates of B-type stars), 
with a secondary criterion involving the presence of at least one harmonic\footnote{While these criteria might appear somewhat generic, more quantitative criteria can only be derived \textit{a posteriori}, as an increasing number of magnetic detections is accomplished, by comparing the characteristics of the magnetic population to the rest of the studied sample.} 
(although the absence of a harmonic does not eliminate a candidate outright, especially if its signal appears roughly sinusoidal and reliably repeatable, as in the case of HD~30963, presented below). 

Given the potential confusion between rotational and orbital signals, especially in the case of ellipsoidal variations, a thorough literature search led to the elimination of known binaries with periods corresponding to 
{those} of the suspected rotational modulation {signals} (or close binaries with unknown periods
{, which should at the very least be followed up with spectroscopy first before being considered for spectropolarimetric follow-up}), as well as known magnetic stars (and stars for which previous spectropolarimetric observations had not yielded a {significant} magnetic detection
). Stars in crowded fields with signals of ambiguous origin were also discarded. 

The final target list contained seven high-probability magnetic candidates
{. Out of these seven stars, four were observed with} follow-up spectropolarimetry 
{from} the Canada-France-Hawaii Telescope (CFHT) 
{during} the 2019B semester. 
These stars are listed in Table~\ref{tab:target_list}, along with their putative {\textit{TESS}} rotational period. Their light curves (detrended following the procedure described by \citealt{2018A&A...616A..77B}), which show similarities with those of known magnetic stars (e.g. \citealt{2019MNRAS.487..304D}), and the associated Lomb-Scargle periodograms are shown in Fig.~\ref{fig:lc}.




\section{Spectropolarimetry}\label{sec:specpol}

\subsection{Observations}

Each target was observed with 
ESPaDOnS (an Echelle SpectroPolarimetric Device for the Observation of Stars; \citealt{2006ASPC..358..362D}) at the CFHT, as part of observing program 19BC34 (PI: David-Uraz). This instrument has a high 
{resolving power} ($R \sim 65000$) and covers a wavelength range of about 3600-10000 \AA. Data reduction was performed using the \textsc{Upena} pipeline \citep{2011tfa..confE..63M}, which is based on the \textsc{Libre-ESpRIT} reduction package \citep{1997MNRAS.291..658D}. It yields integrated spectra (Stokes $I$ parameter), as well as circularly polarized spectra (Stokes $V$ parameter), which are sensitive to the Zeeman effect 
\citep[e.g.][]{2004ASSL..307.....L}, thus allowing us to detect and measure astrophysical magnetic fields
. 

Each {complete} observation consists of 
{four} subexposures, each corresponding to a different angle of the Fresnel rhombs, and their combination can produce each of the aforementioned spectra, as well as two diagnostic nulls {($N$)}, which characterize the level of noise {and allow us to identify potential spurious signals} in the Stokes $V$ spectrum. The number of observations per target and exposure times are detailed in Table~\ref{tab:target_list}, while the 
{longitudinal field measurement} and signal-to-noise ratio 
{(S/N)} achieved 
{for} each observation appear in Table~\ref{tab:app}. It should be noted that for two of the four observations of HD 25709, only two subexposures were obtained, which means that they are not accompanied by a diagnostic null and that the removal of systematics is less effective for these observations.

\begin{table*}
	\centering
	\caption{Table of the targets and observational information. The first two columns provide identifiers, both in the HD and TIC \citep{2019AJ....158..138S} catalogues, {and} the third column lists the \textit{TESS} sectors in which each target was observed. Column 4 lists the measured {fundamental (in cases where harmonics are detected)} photometric period, 
	{assessed} to be of rotational origin (with the uncertainty on the last digit in parentheses). Columns 5-7 list their spectral types, effective temperatures and magnitudes in the $V$ band (with the reference for $T_\textrm{eff}$ provided in the last column). Finally, columns 8 and 9 indicate the number of spectropolarimetric observations {obtained} for each star, as well as the integration time for each subexposure. It should be noted that two {observations} of HD 25709 
	consisted of {only} two subexposures 
	(rather than the usual four), 
	{and can therefore not yield diagnostic nulls for comparison to their Stokes $V$ profiles}.}
	\label{tab:target_list}
	\begin{tabular}{@{}r r c l l c c c r c@{}} 
		\hline
		HD & TIC no. & Sector(s) & Period & Spectral type & $T_\textrm{eff}$ & $V$ & $N_\textrm{obs}$ & $t_\textrm{exp}$ & Reference {($T_\textrm{eff}$)}\\
		   &         &        &    (d) &   & (kK)       & (mag) &  & (s) & \\
		\hline
		25709 & 34199198 & 5 & 2.554(1) & B9V & 9.9 & 7.98 & 4 & 859 & \citet{2012MNRAS.427..343M}\\
		30963 & 9355205 & 5 & 3.9892(9) & B9III & 11.5 & 7.23 & 3 & 318 & \citet{2019AJ....158..157M} \\
		32686 & 213104118 & 5 & 1.2336(4) & B4II/III & 14.2 & 6.03 & 2 & 793 & \citet{2016AnA...591A.118S} \\
		38170 & 140288359 & 5, 6 & 2.76618(4) & B9.5V & 10.3 & 5.28 & 2 & 168 & \citet{2003AnA...408.1065H} \\
		\hline
	\end{tabular}
\end{table*}

\subsection{{Analysis}}\label{sec:results}

For each star, we obtained 
spectral line lists 
using the most up-to-date version of the Vienna Atomic Line Database {(VALD3\footnote{\url{http://vald.astro.uu.se/}}; \citealt{1995A&AS..112..525P,1997BaltA...6..244R,1999A&AS..138..119K,2000BaltA...9..590K,2015PhyS...90e4005R})}. Details about the abundances and stellar parameters used are 
{in the following subsections}. Using dedicated 
\textsc{idl} routines, we then normalized the spectra {to the continuum} and constructed line masks to perform Least-Squares Deconvolution (LSD; \citealt{1997MNRAS.291..658D}), a multi-line technique that increases the 
{S/N} of the Stokes $V$ signatures upon which our magnetometric analysis is computed. Hydrogen lines were excluded, 
as they violate the assumption of line self-similarity required by the LSD algorithm
, as well as {lines} severely blended 
{with hydrogen} and lines contaminated by telluric absorption. 

Once the LSD profiles were computed {using the iLSD 
routine \citep{2010A&A...524A...5K}}, we calculated the disc-averaged longitudinal field 
using the first-order moment method (e.g. \citealt{2000MNRAS.313..851W}) and we also calculated false alarm probabilities {(FAPs)} as discussed by \citet{1997MNRAS.291..658D}: a FAP {computed on the Stokes $V$ profile within the bounds of the LSD profile} of less than 10$^{-5}$ is considered to correspond to a definite detection (DD), while a FAP between 10$^{-3}$ and 10$^{-5}$ constitutes a marginal detection (MD); anything higher is considered a non-detection (ND). Longitudinal field measurements and results of the FAP determinations are presented in Table~\ref{tab:app}.

Finally, we also perform a Bayesian analysis (as described by \citealt{2012MNRAS.420..773P}), {using unconstrained rotational phases}, to determine 
{the surface} field strengths {that} are compatible with the obtained LSD Stokes $V$ profiles assuming an oblique global dipolar field configuration \citep{1950MNRAS.110..395S} and marginalizing\footnote{In the context of Bayesian statistics, this refers to the procedure of integrating over ``nuisance" parameters to obtain a \textit{marginal} distribution with respect to the parameter(s) of interest -- in this case, dipolar field strength.} the probability density function over the geometric parameters (inclination and obliquity). The results are presented in each individual subsection below.

\begin{table}
	\centering
	\caption{Observing log (spectropolarimetric data). For each observation, the HD number, date (in the format MHJD $\equiv$ HJD - 2,450,000), and observation identifier are  provided (columns 1, 2 and 3) as well as the peak signal-to-noise ratio 
	{(S/N)} in the Stokes $I$ spectrum, per spectral pixel (column 6). The computed longitudinal field measurement ($B_\textrm{z}$) and detection status (``DET'', see Section~\ref{sec:results}; N corresponds to a non-detection while {M denotes a marginal detection and} D corresponds to a definite detection) are indicated in columns 4 and 5, respectively. An asterisk next to the HD number corresponds to an observation for which only 2 subexposures were obtained; thus these observations do not have a diagnostic null. Finally, the results for HD 25709 are designated as HD 25709A (for the primary in our binary analysis) and HD 25709B (for the secondary), with the longitudinal field being measured from disentangled LSD profiles.}
	\label{tab:app}
	\begin{tabular}{@{}l c c r@{} l c r@{}} 
		\hline
		HD & MHJD & ObsID & \multicolumn{2}{c}{$B_\textrm{z}$} & DET? & {S/N} \\
		   & & & \multicolumn{2}{c}{(G)} &  & \\
		\hline
		25709A & 8744.05269 & 2437131p & 69$\pm$ & 51 & N & 552\\
		25709A* & 8804.91663 & 2460018p & -145$\pm$ & 161 & N & 248\\
		25709A* & 8806.92018 & 2460205p & 5$\pm$ & 178 & M & 223\\
		25709A & 8806.98866 & 2460207p & 24$\pm$ & 133 & N & 223\\
		\hline
		25709B & 8744.05269 & 2437131p & 0$\pm$ & 29 & N & 552\\
		25709B* & 8804.91663 & 2460018p & 12$\pm$ & 83 & N & 248\\
		25709B* & 8806.92018 & 2460205p & -46$\pm$ & 101 & N & 223\\
		25709B & 8806.98866 & 2460207p & 192$\pm$ & 75 & N & 223\\
		\hline
		30963 & 8743.13423 & 2436982p & 5$\pm$ & 16 & N & 461 \\
		30963 & 8807.01889 & 2460211p & -13$\pm$ & 28 & N & 271 \\
		30963 & 8807.03561 & 2460215p & -16$\pm$ & 23 & N & 327 \\
		\hline
		32686 & 8744.09261 & 2437135p & 21$\pm$ & 13 & N & 1267 \\
		32686 & 8807.06450 & 2460219p & -39$\pm$ & 21 & N & 837\\
		\hline
		38170 & 8741.14509 & 2436777p & -6$\pm$&25 & D & 458 \\
		38170 & 8743.14585 & 2436986p & 105$\pm$&14 & D & 685 \\
		\hline
	\end{tabular}
\end{table}

\subsection{{Results}}

\subsubsection{HD 25709}

\begin{figure}
	\includegraphics[width=\columnwidth]{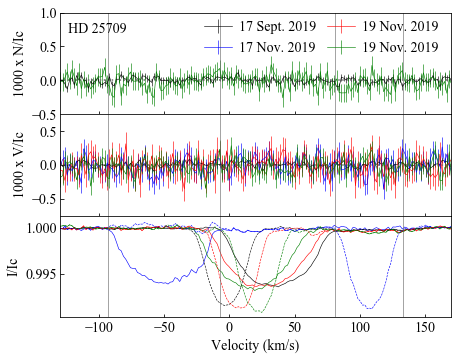}
    \caption{
    {Disentangled LSD {Stokes $I$ (bottom)} profiles of both components of HD 25709 (the primary is shown with a 
    {solid} line, and the secondary with a dashed line)
    , {shown with the non-disentangled Stokes} $V$ {profiles} (middle) and the diagnostic nulls (top), for four different (colour-coded) 
    {observations} (except for two diagnostic nulls which were not available). Thin vertical lines correspond to the integration ranges used for magnetometric analysis, here represented for the best separated profiles for each component. All Stokes $V$ profiles are consistent with non-detections, while the Stokes $I$ profiles obviously show the SB2 nature of this system.}}
    \label{fig:lsd25709}
\end{figure}

\begin{figure}
	\includegraphics[width=\linewidth]{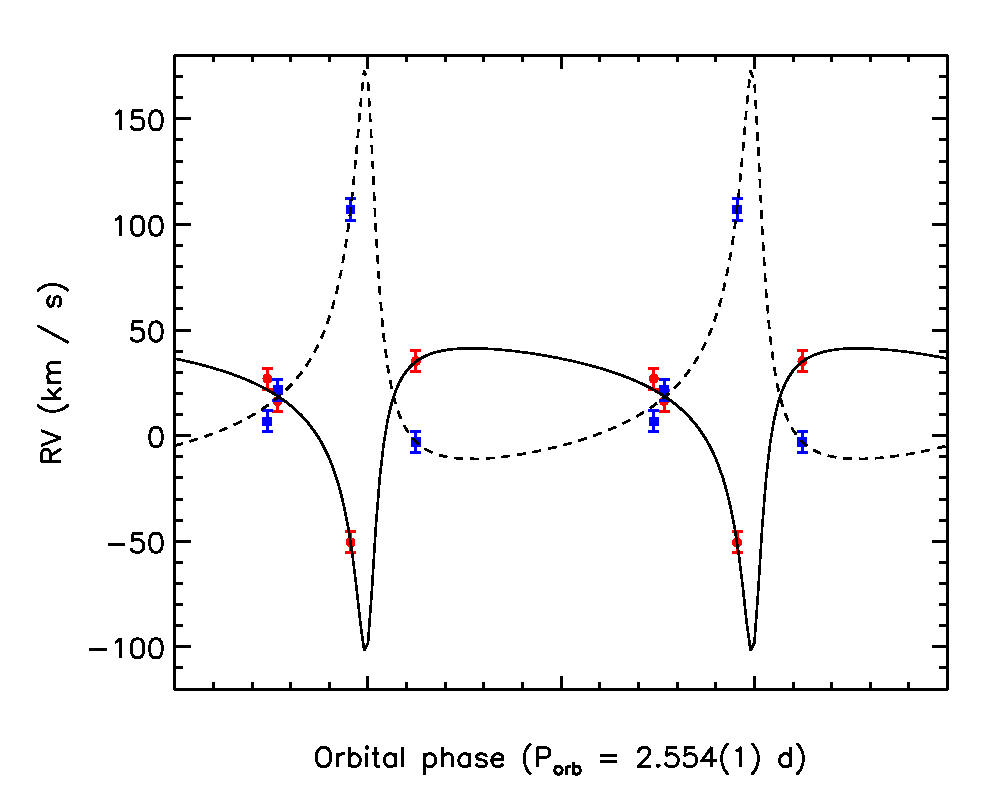}
	\includegraphics[width=\linewidth]{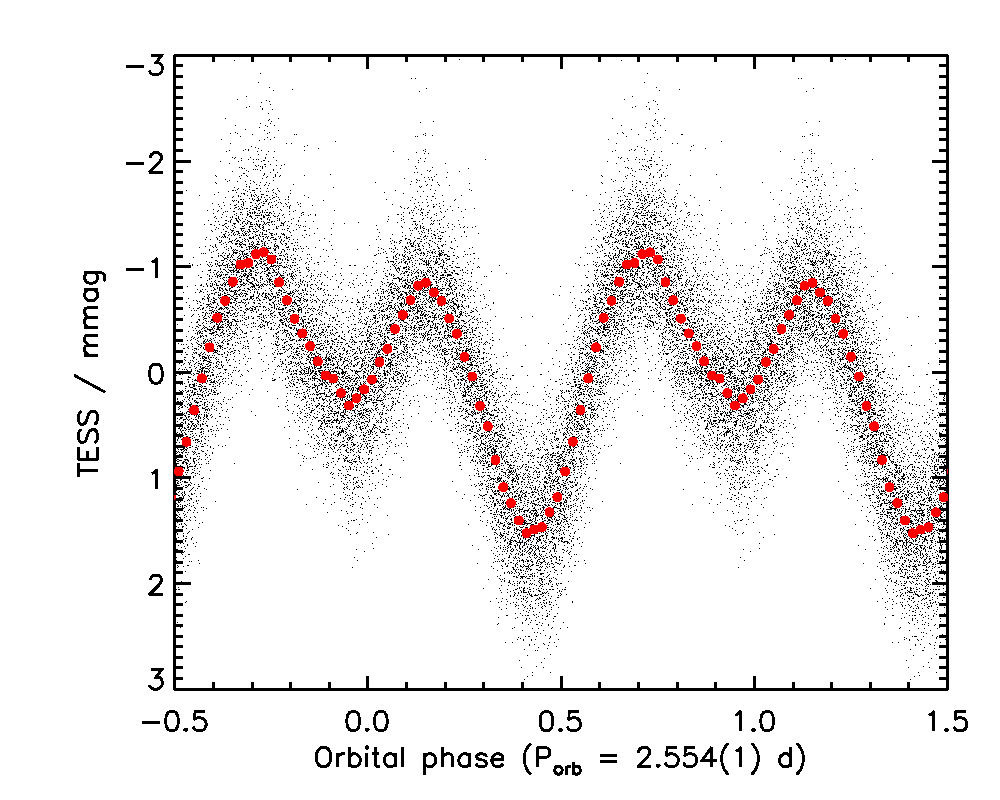}
    \caption{Phased radial velocity (RV) curve (top) and light curve (bottom). In the top panel, RV measurements of the primary are shown in red (with the best fit represented by a solid black line), while measurements of the secondary are shown in blue (with a dashed curve to represent the best fit). In the bottom panel, red dots represent observations binned in phase.}
    \label{fig:binary}
\end{figure}

HD 25709 is a 
poorly studied object, with very few mentions in the literature. Classified as a B9V star by \citet{1999MSS...C05....0H}, 
{an} examination of the ESPaDOnS spectra 
reveals it to 
be an SB2 binary {(see Fig.~\ref{fig:lsd25709})}. Using a 
line list computed assuming {solar} abundances and the published values of $T_\textrm{eff}$ and $\log g$ \citep{2012MNRAS.427..343M}, we extracted LSD profiles for each observation {with a single mask}, 
fitting the lines to measure radial velocities and projected rotational velocities ($v_\textrm{A} \sin i_\textrm{A}$ = 39.0$\pm$0.8 km$\,$s$^{-1}$, $v_\textrm{B} \sin i_\textrm{B}$ = 19.8$\pm$0.2 km$\,$s$^{-1}$) for both components\footnote{Although the line mask used to generate the LSD profiles is most likely more appropriate for the primary star than it is for the secondary, the difference in temperature between both components is not large enough for this to affect the radial velocity measurements, since they both clearly appear in the fitted LSD profiles.}. We then fit a Keplerian orbit to the radial velocities, assuming that the inferred \textit{TESS} photometric period is orbital in nature, using a Markov Chain Monte-Carlo (MCMC) sampler. The resulting radial velocity curve {and its best fit}, as well as the associated phased {\textit{TESS}} light curve, are presented in 
{Fig.~\ref{fig:binary}}.

This allows us to derive orbital and physical parameters (for a full list of the derived parameters, see Table~\ref{tab:appen}) {for the system given the aforementioned assumption that the orbital period is known {from \textit{TESS} photometry} ($P_\textrm{orb}$ = 2.554 d)}
; 
we find a particularly eccentric system {($e$ = 0.73$\pm$0.01) with a low projected total mass ($(M_\textrm{A}+M_\textrm{B})\sin^{3} i_\textrm{orb}$ =  $0.37\pm0.03$ M$_\odot$).}
{Disentangling was performed on the Stokes $I$ LSD profiles through an iterative process (Fig.~\ref{fig:lsd25709}; \citealt{2006A&A...448..283G}, as implemented by \citealt{2018MNRAS.475..839S, 2018MNRAS.475.5144S}). Stokes $V$ was not disentangled, but the integration ranges for the magnetometric analysis were based on the disentangled Stokes $I$ profiles for each component, allowing us to perform a separate analysis on each component of the system.}

Given the published values of parallax ($\pi$ = 3.36$\pm$0.06 mas; \citealt{2018A&A...616A...1G}), the extinction value ($E(B-V) = 0.01$) based on a comparison between the observed photometry and an interpolation of the intrinsic colours derived by \citet{2013ApJS..208....9P} for a $T_\textrm{eff}$ of 9900 K, and the associated bolometric correction ($BC = -0.28$), we find the luminosity of the system to be consistent with late main sequence B-type stars. Combined with the low projected total mass, this leads to the conclusion that the inclination $i$ is likely low, therefore the photometric variations are 
{unlikely to be} caused by eclipses. 
Furthermore, given the high inferred eccentricity, this system would be a good candidate to exhibit ``heartbeat'' variations (though they present varied morphologies, these variations, which are due to dynamic tidal distortions in eccentric binary systems, tend to be reminiscent of an echocardiogram; \citealt{2012ApJ...753...86T}); however, the shape of the light curve does not correspond to the pattern of variability that would be expected in such a case. 
{On the other hand,} the photometric period 
{may not be} orbital after all; it is possible that the system possesses a different orbital period (which cannot currently be constrained using only four radial velocity measurements{, though it must be relatively short since the radial velocity of the secondary varies by $\sim$100 km\,s$^{-1}$ over two nights}), and that the photometric signal is in fact linked to rotational modulation in one of the stars. Further observations {(especially spectroscopy to extract new radial velocity measurements)} 
{are} required to better understand this system
.

As for the magnetometric analysis, all disentangled profiles 
{led} to non-detections {(with a smallest {1$\sigma$} error bar on $B_\textrm{z}$ of 51 G for the primary and 29 G for the secondary)} 
{with one exception} (for which FAP = $1.92 \times 10^{-4}$). While this would normally be considered as a marginal detection, we note that it occurs in one of the observations for which there is no diagnostic null. Therefore, this result should be viewed as inconclusive. 

Finally, applying the Bayesian analysis of \citet{2012MNRAS.420..773P}
{ to the two observations consisting of full spectropolarimetric sequences (four sub-exposures)}, we find upper limits {on the polar strength of the dipole field ($B_\textrm{d, max}$)} of {408~G} for the primary and {244~G} for the secondary 
{(corresponding to the 95.4 per cent credible region of the probability density function, marginalized for the field strength).}

\subsubsection{HD 30963}

\begin{figure}
	\includegraphics[width=\columnwidth]{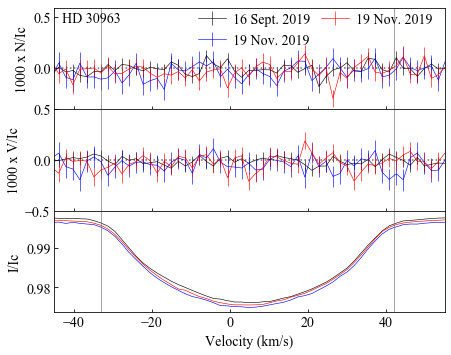}
    \caption{{LSD profiles of HD 30963 in Stokes $I$ (bottom), $V$ (middle) and the diagnostic nulls (top), for three different observations (colour-coded). The integration range used for magnetometric analysis is represented by thin vertical lines. All Stokes $V$ profiles are consistent with non-detections.}}
    \label{fig:lsd30963}
\end{figure}

A late B-type star, HD 30963 has not been extensively studied in the past, and was not known to be chemically peculiar at the time of our spectropolarimetric follow-up proposal. In the intervening time however, an abundance analysis based on high-resolution spectra revealed it to be a HgMn star \citep{2019AJ....158..157M}. So far, no strong, globally organized magnetic field has yet been detected at the surface of such an object \citep{2013A&A...554A..61K}. 
We computed a line mask using the abundances and atmospheric parameters of \citet{2019AJ....158..157M} and performed our magnetometric analysis {(our LSD profiles are shown in Fig.~\ref{fig:lsd30963})}. This resulted in non-detections ({with a smallest error bar on the longitudinal field of 16 G; }Table~\ref{tab:app}). 
{Our Bayesian analysis yields in turn} an upper limit of $B_\textrm{d, max}$ = {63~G} {(95.4 per cent credible region)}. {Nevertheless, we evaluate that it is probable that the signal recovered by \textit{TESS} is rotational in origin, as has been observed in similar objects \citep[e.g.][]{2020MNRAS.492.1834P}.}

\subsubsection{HD 32686}

\begin{figure}
	\includegraphics[width=\columnwidth]{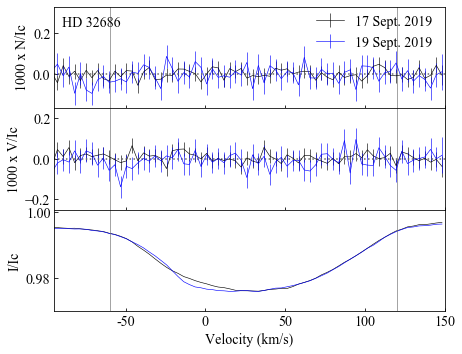}
    \caption{Same as {Fig.~\ref{fig:lsd30963}}, but for HD 32686; note that there appear to be slight variations in the Stokes $I$ line profile.}
    \label{fig:lsd32686}
\end{figure}

Classified as a B4II/III by \citet{1999MSS...C05....0H}, HD 32686 is a fairly evolved star presumably approaching the end of its {main sequence} lifetime. It has not previously been noted to exhibit chemical peculiarities{, and it should be noted that it does not show an excess in $\Delta a$ photometry\footnote{This photometric index \citep{1976A&A....51..223M} leverages the flux depression in magnetic chemically peculiar stars at about 5200 \AA, and its efficiency to reliably identify such stars was confirmed in several studies (e.g. \citealt{1983A&A...123...48M}).} \citep{1998A&AS..130..455V}}, therefore we adopted {solar} abundances, together with the atmospheric parameters tabulated by \citet{2016AnA...591A.118S}, to compute a VALD3 line list to use in our magnetometric analysis. This resulted in non-detections{, with a best longitudinal field error bar of 23 G}. However, apparent line profile variations {(Fig.~\ref{fig:lsd32686})} between the two observations, combined with 
seemingly strong helium lines{,} led {us} to the suspicion that this might be a chemically peculiar star nonetheless. 
Consequently, we tested a line mask {previously} computed for a Bp star of similar effective temperature {(HD 145501C; \citealt{2017MNRAS.468.2745N})}, which yielded more stringent constraints on the longitudinal field {due to the increased number of lines}, lowering the error bar to 13 G. These results are the ones presented in Table~\ref{tab:app} (and they are still non-detections). We also based our Bayesian analysis on the LSD profiles yielded by the second mask, which leads to an upper limit of $B_\textrm{d, max}$ = {140~G} {(95.4 per cent credible region)}
.

\subsubsection{HD 38170}

Although it is not listed as being chemically peculiar in the \citet{2009A&A...498..961R} catalog, HD 38170 (= WZ Col) was found to be overabundant in strontium and barium by \citet{2003AnA...408.1065H} and identified as a probable $\alpha^2$ CVn variable by \citet{2011MNRAS.414.2602D} based on Hipparcos 
{photometry ($P$ = 1.38 d; we infer a period twice as long from the \textit{TESS} light curve)}. {It also shows a strong excess in $\Delta a$ photometry \citep{1998A&AS..130..455V}, further 
suggesting that it is 
a chemically peculiar star.} 
{Moreover, as} the brightest star in our sample, HD 38170 thus represented \textit{a priori} the most promising candidate to detect a magnetic field among the four objects which form this study.

\begin{figure}
	\includegraphics[width=\columnwidth]{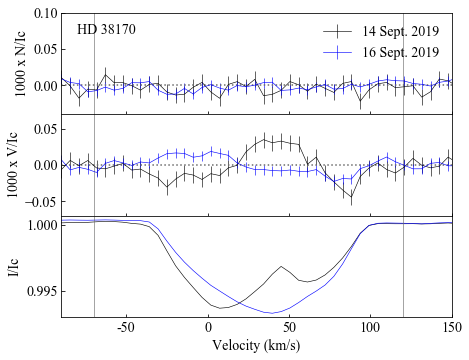}
    \caption{
    Same as {Fig.~\ref{fig:lsd30963}}, but for HD 38170. Notice the asymmetry in the black Stokes $I$ profile, consistent with the presence of chemical spots on the surface of the star, as seen in, e.g., HR 2949 \citep{2015MNRAS.449.3945S}. Both Stokes $V$ profiles are consistent with a definite magnetic detection.}
    \label{fig:lsd}
\end{figure}

We constructed our line mask using the abundances derived by \citet{2003AnA...408.1065H}, as well as {non-solar} abundances for chromium and rare-Earth elements, based on their apparent temperature dependence in ApBp stars\footnote{We used nominal values of log([Cr/H]) = -4.0, log([Pr/H]) = log([Nd/H]) = log([Eu/H]) = log([Ce/H]) = -8.0.} \citep{2017AstL...43..252R}. Both observations yielded a definite detection {(based on the FAPs)} and the longitudinal field measured from the second observation 
is non-null at the 7.5$\sigma$ 
level ($B_\textrm{z}$ = 105$\pm$14 G). The associated LSD profiles are shown in Fig.~\ref{fig:lsd}, and a clear signal can be seen in Stokes $V$. As a result, we conclude that this star constitutes a firm new magnetic detection.

Using the Gaia parallax ($\pi$ = {8.90$\pm$0.15 mas}; \citealt{2018A&A...616A...1G}), expected intrinsic colour ($B-V = -0.04$, based on the value of $T_\textrm{eff}$ provided in Table~\ref{tab:target_list}; \citealt{2013ApJS..208....9P}) and bolometric correction (accounting for its chemical peculiarity, $BC = -0.21$; \citealt{2008A&A...491..545N}) for this star, we derive, based on the $V$ magnitude listed in Table~\ref{tab:target_list}, its stellar parameters using the MCMC sampler described by \citet{2019MNRAS.485.1508S}. The results are listed in Table~\ref{tab:atm}. Based on the obtained radius and the measured value of $v \sin i$ {(57$\pm$5 km\,s$^{-1}$)}, as well as the period 
{determined from} the \textit{TESS} light curve (assumed to be rotationally modulated), we can also constrain the inclination $i$
. 
{It is} 
{included} in Table~\ref{tab:atm}.

\begin{table}
	\centering
	\caption{Stellar 
	parameters of HD 38170, fit with an MCMC sampler, in order: luminosity, 
	surface gravity (in cgs units), polar radius, mass, age {(based on the evolutionary tracks of \citealt{2012A&A...537A.146E})}, fractional main-sequence age 
	{and inclination of the rotational axis with respect to the line of sight.}
	}
	\label{tab:atm}
	\begin{tabular}{l c} 
		\hline
		\multicolumn{2}{c}{Stellar parameters}\\
		\hline
		$\log (L/\textrm{L}_\odot)$ & {$2.0\pm0.1$} \rule[-1.4ex]{0pt}{0pt}\\
		$\log g$ & {$3.86\pm0.06$} \rule[-1.4ex]{0pt}{0pt}\\
		$R_\textrm{p}$ (R$_\odot$) &  {$3.3\pm0.3$} \rule[-1.4ex]{0pt}{0pt}\\
		$M_*$ (M$_\odot$) & {$2.8\pm0.1$} \rule[-1.4ex]{0pt}{0pt}\\
		Age (Myr) & {$394^{+10}_{-17}$} \rule[-1.4ex]{0pt}{0pt}\\
		$t/t_\textrm{TAMS}$ & {$0.68^{+0.07}_{-0.04}$} \rule[-1.4ex]{0pt}{0pt}\\
        {$i$ (\textdegree)} & {67$\pm$6}\\
		\hline
	\end{tabular}
\end{table}

Finally, 
we 
{performed} a 
Bayesian analysis on our LSD profiles, yielding a dipolar field strength of $B_\textrm{d}$ = {$\textrm{254}^{+\textrm{78}}_{-\textrm{49}}$~G} 
(with a 
{68.3 per cent} {credible region}
)
, based on the probability density function normalized over the geometric parameters ($i$ and $\beta$). 
{We note that our modelling assumes 
the same intensity line profile for all observations. Therefore, the line profile variability that is evident in the Stokes $I$ profiles of HD 38170 is not taken into account, but we consider this to be a reasonable approach given the fact that we reproduce the general shape of the Stokes $V$ profiles satisfactorily nonetheless.}

{The fairly weak dipolar field strength that we obtain }is consistent with the moderately high fractional main-sequence age, assuming flux conservation. {The model with the maximum \textit{a posteriori} probability yielded values of $B_\textrm{d}$ = 257 G, $i$ = 57\textdegree and $\beta$ = 60\textdegree, with a difference in rotational phase between both observations of $\Delta \phi$ = 0.7222, which is almost identical to what we obtain given the \textit{TESS} photometric period (0.7233). We note that the value of the inclination falls just a bit outside of the range obtained from the stellar parameter analysis; marginalizing the two-dimensional posterior probability density function between the inclination and obliquity for values of $i$ located within that range, we find that $\beta$ = $45^{+10}_{-5}$\,\textdegree
. While this preliminary work seems to indicate that a dipolar model can explain 
the essential characteristics of our observations
, more observations and sophisticated modelling (such as Zeeman Doppler Imaging or ZDI; \citealt{1989A&A...225..456S}) would be required for a full magnetic characterization of this star.} 

\section{
Summary and discussion}\label{sec:conclusions}

Out of four {B-type} stars that were selected photometrically {to exhibit rotational modulation from the \textit{TESS} mission}, one was found to host a detectable magnetic field. We placed upper limits on the magnetic field strength inferred from the non-detections. A previously unknown binary system (HD 25709) was also found
, although some ambiguity remains regarding its orbital solution and the nature of its photometric variations. The detection was achieved for HD~38170, which apparently is an evolved late B-type star. The fairly weak inferred field strength ($B_\textrm{d} \sim$ {250} G) is compatible with its proposed evolutionary status, and therefore this new detection 
{lies} in an undersampled region of the Hertzsprung-Russell diagram {\citep[e.g.][]{2020pase.conf..170P}}, given the dearth of known evolved magnetic massive stars {\citep{2016A&A...592A..84F}}. A more detailed characterization of its field geometry would require follow-up observations and a more detailed analysis. Given its inferred age, this star might offer insights into the evolution of the strength and complexity of magnetic fields in OBA stars over their main-sequence lifetimes. 

Of the remaining two stars, HD~30963 is an HgMn star, and therefore the non-detection of a magnetic field on its surface falls in line with the lack of detection of strong large-scale fields in this class of stars (e.g. \citealt{2013A&A...554A..61K}). Nevertheless, the photometric variability observed in its \textit{TESS} light-curve still might be rotational in nature, as abundance spots on the surface of the star may lead to rotational modulation of its brightness, even though the periodicity might not remain over large time scales \citep{2020MNRAS.492.1834P}. As for HD~32686, a similar result might not be entirely surprising given its apparent evolutionary 
{stage}. Under the assumption of flux conservation, typical dipolar fields observed in {main sequence} 
stars might weaken 
by about an order of magnitude over their main-sequence lifetimes (e.g. \citealt{2019MNRAS.485.5843K}), potentially placing their strength near the terminal age main sequence just below the magnetic sensitivity achieved in this study. Interestingly, this star also exhibits the weakest photometric variations 
of the four stars 
{that} were observed. 
{This could potentially hint at the presence of a very weak field on the surface of HD 32686, below our current detection limits, although there is no demonstrated correlation between field strength and the amplitude of rotationally modulated light-curve variability in magnetic B-type stars.} {That said, its relatively short rotation period ($\sim$ 1.23 d) is somewhat surprising in that regard, as a magnetic star would surely have undergone a significant amount of magnetic braking over its lifetime \citep[e.g.][]{2009MNRAS.392.1022U}.}

While our detection rate of 25 
{per cent} 
{is significantly higher than the observed incidence of magnetic OBA stars, 
{it is} still lower than the detection rate that we aim to achieve with our larger targeted survey ($> 50$ per cent}){. However}, we do not consider it to be representative of the potential of \textit{TESS} data to select magnetic candidates, because of the extremely small number statistics. That said, 
we consider this initial effort to provide a satisfactory proof of concept of the method we are proposing to use: selecting high-probability candidate magnetic stars photometrically to perform high-yield spectropolarimetric surveys. Stars that are detected to host a surface magnetic field can then be further studied in follow-up investigations to better constrain their stellar and magnetic properties, providing invaluable clues to better understand the origin and evolution of magnetism in massive stars. In the sample presented in this study, even stars that did not end up yielding magnetic detections still showed spectroscopic variability, offering potential clues about non-magnetic sources of photometric variability.

The results of this preliminary study 
also (re)emphasize certain criteria that must be applied, beyond the possible detection of rotational modulation in a light curve (since not all rotational -- or seemingly rotational -- signals are associated to magnetism, as evidenced by this study), to assemble an appropriate target list for such a high-yield survey. In particular, the following types of targets should not be considered for spectropolarimetric follow-up in the context of our targeted survey (although they remain interesting in their own right):

\begin{itemize}
    \item binary (or multiple) stars with orbital periods compatible with the periods extracted from the photometry, or with unknown orbital periods{; we would argue against removing binaries with a known orbital period that is different from the one that is recovered from their light curve, as magnetic OBA stars in binaries (and especially close binaries) are of particular interest \citep{2015IAUS..307..330A}
    }, especially given the insights that they might provide regarding the origin of stellar magnetism in the upper H-R diagram;
    \item HgMn stars, which systematically do not show 
    {evidence} of strong, globally organized fields {despite occasionally exhibiting rotationally modulated photometric signals}, as well as classical Be stars, for which no magnetic field has been detected to date \citep{2016ASPC..506..207W}, as such a field would potentially disrupt a {Keplerian} disc \citep{2018MNRAS.478.3049U};
    \item objects whose light {curves} only {show} very low amplitude variations (such as HD 32686), since although these variations might still be due to magnetism, there is a higher likelihood overall that they might be caused by, e.g., contamination in a crowded field than in cases with higher amplitudes 
    (although this criterion should perhaps not be applied to O stars, as magnetospheric scattering 
    {might lead} to smaller variations than photospheric spots{; \citealt{2020MNRAS.492.1199M}}).
\end{itemize}

Despite not being primary targets for the purposes of our targeted survey, it should be noted that dedicated efforts to achieve high precision magnetometric measurements of HgMn stars remain warranted, as the origin of their chemical peculiarities is still unknown and hypothesized to be related to magnetism.

Finally, while the number of candidate magnetic stars involved in this effort precludes an efficient use of machine learning methods, future detections will nonetheless help us {to} iteratively improve our candidate selection criteria
{. This will lead} to a high-yield survey and ultimately better statistics for the population of massive and intermediate-mass magnetic stars.

\section*{Acknowledgements}

This paper includes data collected by the \textit{TESS} mission. Funding for the \textit{TESS} mission is provided by the NASA Explorer Program. Funding for the \textit{TESS} Asteroseismic Science Operations Centre is provided by the Danish National Research Foundation (Grant agreement no.: DNRF106), ESA PRODEX (PEA 4000119301) and Stellar Astrophysics Centre (SAC) at Aarhus University. We thank the \textit{TESS} and TASC/TASOC teams for their support of the present work. This research has made use of the SIMBAD database, operated at CDS, Strasbourg, France. Some of the data presented in this paper were obtained from the Mikulski Archive for Space Telescopes (MAST). STScI is operated by the Association of Universities for Research in Astronomy, Inc., under NASA contract NAS5-2655. This work has made use of the VALD database, operated at Uppsala University, the Institute of Astronomy RAS in Moscow, and the University of Vienna.

ADU {and GAW acknowledge} 
the support of the Natural Science and Engineering Research Council of Canada (NSERC). This work is supported by NASA under award number 80GSFC17M0002. MES acknowledges the financial support provided by the Annie Jump Cannon Fellowship, supported by the University of Delaware and endowed by the Mount Cuba Astronomical Observatory. {VP acknowledges support from the National Science Foundation under Grant No. 1747658.} The research leading to these results has received funding from the European Research Council (ERC) under the European Union's Horizon 2020 research and innovation programme (grant agreement No. 670519: MAMSIE), and a senior postdoctoral fellowship from the Research Foundation Flanders (FWO) with grant agreement No. 1286521N to DMB.

{Finally, the authors also thank Z. Keszthelyi, J. Krti\v{c}ka, A.~F.~J. Moffat, E. Paunzen and E. Semenko for their helpful comments, as well as the referee, E. Alecian, for contributing to the improvement of this paper.}

\section*{Data Availability}

The spectropolarimetric data underlying this article are available in the Canadian Astronomy Data Center (CADC) archive at \url{https://www.cadc-ccda.hia-iha.nrc-cnrc.gc.ca/en/} 
and are uniquely identified with the ObsIDs listed in Table~\ref{tab:app}.




\bibliographystyle{mnras}
\bibliography{refs} 

\begin{thebibliography}{}
\makeatletter
\relax
\def\mn@urlcharsother{\let\do\@makeother \do\$\do\&\do\#\do\^\do\_\do\%\do\~}
\def\mn@doi{\begingroup\mn@urlcharsother \@ifnextchar [ {\mn@doi@}
  {\mn@doi@[]}}
\def\mn@doi@[#1]#2{\def\@tempa{#1}\ifx\@tempa\@empty \href
  {http://dx.doi.org/#2} {doi:#2}\else \href {http://dx.doi.org/#2} {#1}\fi
  \endgroup}
\def\mn@eprint#1#2{\mn@eprint@#1:#2::\@nil}
\def\mn@eprint@arXiv#1{\href {http://arxiv.org/abs/#1} {{\tt arXiv:#1}}}
\def\mn@eprint@dblp#1{\href {http://dblp.uni-trier.de/rec/bibtex/#1.xml}
  {dblp:#1}}
\def\mn@eprint@#1:#2:#3:#4\@nil{\def\@tempa {#1}\def\@tempb {#2}\def\@tempc
  {#3}\ifx \@tempc \@empty \let \@tempc \@tempb \let \@tempb \@tempa \fi \ifx
  \@tempb \@empty \def\@tempb {arXiv}\fi \@ifundefined
  {mn@eprint@\@tempb}{\@tempb:\@tempc}{\expandafter \expandafter \csname
  mn@eprint@\@tempb\endcsname \expandafter{\@tempc}}}

\bibitem[\protect\citeauthoryear{{Alecian} et~al.,}{{Alecian}
  et~al.}{2015}]{2015IAUS..307..330A}
{Alecian} E.,  et~al., 2015, in {Meynet} G.,  {Georgy} C.,  {Groh} J.,   {Stee}
  P.,  eds,  IAU Symposium Vol. 307, New Windows on Massive Stars. pp 330--335
  (\mn@eprint {arXiv} {1409.1094}), \mn@doi{10.1017/S1743921314007030}

\bibitem[\protect\citeauthoryear{{Babcock}}{{Babcock}}{1958}]{1958ApJ...128..228B}
{Babcock} H.~W.,  1958, \mn@doi [\apj] {10.1086/146539}, \href
  {https://ui.adsabs.harvard.edu/abs/1958ApJ...128..228B} {128, 228}

\bibitem[\protect\citeauthoryear{{Bagnulo} et~al.,}{{Bagnulo}
  et~al.}{2020}]{2020A&A...635A.163B}
{Bagnulo} S.,  et~al., 2020, \mn@doi [\aap] {10.1051/0004-6361/201937098},
  \href {https://ui.adsabs.harvard.edu/abs/2020A&A...635A.163B} {635, A163}

\bibitem[\protect\citeauthoryear{{Balona}}{{Balona}}{2013}]{2013MNRAS.431.2240B}
{Balona} L.~A.,  2013, \mn@doi [\mnras] {10.1093/mnras/stt322}, \href
  {https://ui.adsabs.harvard.edu/abs/2013MNRAS.431.2240B} {431, 2240}

\bibitem[\protect\citeauthoryear{{Barron}, {Wade}, {Bowman}, {David-Uraz},
  {Munoz}, {Pablo}  \& {Sim{\'o}n-D{\'\i}z}}{{Barron}
  et~al.}{2020}]{2020pase.conf..226B}
{Barron} J.,  {Wade} G.~A.,  {Bowman} D.~M.,  {David-Uraz} A.,  {Munoz} M.~S.,
  {Pablo} H.,   {Sim{\'o}n-D{\'\i}z} S.,  2020, in {Wade} G.,  {Alecian} E.,
  {Bohlender} D.,   {Sigut} A.,  eds, ~ Vol. 11, Stellar Magnetism: A Workshop
  in Honour of the Career and Contributions of John D. Landstreet. pp 226--235
  (\mn@eprint {arXiv} {2001.04534})

\bibitem[\protect\citeauthoryear{{Bowman}, {Buysschaert}, {Neiner},
  {P{\'a}pics}, {Oksala}  \& {Aerts}}{{Bowman}
  et~al.}{2018}]{2018A&A...616A..77B}
{Bowman} D.~M.,  {Buysschaert} B.,  {Neiner} C.,  {P{\'a}pics} P.~I.,  {Oksala}
  M.~E.,   {Aerts} C.,  2018, \mn@doi [\aap] {10.1051/0004-6361/201833037},
  \href {http://adsabs.harvard.edu/abs/2018A%26A...616A..77B} {616, A77}

\bibitem[\protect\citeauthoryear{{Buysschaert}, {Neiner}, {Martin}, {Aerts},
  {Bowman}, {Oksala}  \& {Van Reeth}}{{Buysschaert}
  et~al.}{2018}]{2018MNRAS.478.2777B}
{Buysschaert} B.,  {Neiner} C.,  {Martin} A.~J.,  {Aerts} C.,  {Bowman} D.~M.,
  {Oksala} M.~E.,   {Van Reeth} T.,  2018, \mn@doi [\mnras]
  {10.1093/mnras/sty1190}, \href
  {https://ui.adsabs.harvard.edu/abs/2018MNRAS.478.2777B} {478, 2777}

\bibitem[\protect\citeauthoryear{{David-Uraz} et~al.,}{{David-Uraz}
  et~al.}{2019}]{2019MNRAS.487..304D}
{David-Uraz} A.,  et~al., 2019, \mn@doi [\mnras] {10.1093/mnras/stz1181}, \href
  {https://ui.adsabs.harvard.edu/abs/2019MNRAS.487..304D} {487, 304}

\bibitem[\protect\citeauthoryear{{Donati} \& {Landstreet}}{{Donati} \&
  {Landstreet}}{2009}]{2009ARA&A..47..333D}
{Donati} J.~F.,  {Landstreet} J.~D.,  2009, \mn@doi [\araa]
  {10.1146/annurev-astro-082708-101833}, \href
  {https://ui.adsabs.harvard.edu/abs/2009ARA&A..47..333D} {47, 333}

\bibitem[\protect\citeauthoryear{{Donati}, {Semel}, {Carter}, {Rees}  \&
  {Collier Cameron}}{{Donati} et~al.}{1997}]{1997MNRAS.291..658D}
{Donati} J.~F.,  {Semel} M.,  {Carter} B.~D.,  {Rees} D.~E.,   {Collier
  Cameron} A.,  1997, \mn@doi [\mnras] {10.1093/mnras/291.4.658}, \href
  {https://ui.adsabs.harvard.edu/abs/1997MNRAS.291..658D} {291, 658}

\bibitem[\protect\citeauthoryear{{Donati}, {Catala}, {Landstreet}  \&
  {Petit}}{{Donati} et~al.}{2006}]{2006ASPC..358..362D}
{Donati} J.~F.,  {Catala} C.,  {Landstreet} J.~D.,   {Petit} P.,  2006,
  {ESPaDOnS: The New Generation Stellar Spectro-Polarimeter. Performances and
  First Results}.
p.~362

\bibitem[\protect\citeauthoryear{{Dubath} et~al.,}{{Dubath}
  et~al.}{2011}]{2011MNRAS.414.2602D}
{Dubath} P.,  et~al., 2011, \mn@doi [\mnras]
  {10.1111/j.1365-2966.2011.18575.x}, \href
  {https://ui.adsabs.harvard.edu/abs/2011MNRAS.414.2602D} {414, 2602}

\bibitem[\protect\citeauthoryear{{Eikenberry} et~al.,}{{Eikenberry}
  et~al.}{2014}]{2014ApJ...784L..30E}
{Eikenberry} S.~S.,  et~al., 2014, \mn@doi [\apjl]
  {10.1088/2041-8205/784/2/L30}, \href
  {https://ui.adsabs.harvard.edu/abs/2014ApJ...784L..30E} {784, L30}

\bibitem[\protect\citeauthoryear{{Ekstr{\"o}m} et~al.,}{{Ekstr{\"o}m}
  et~al.}{2012}]{2012A&A...537A.146E}
{Ekstr{\"o}m} S.,  et~al., 2012, \mn@doi [\aap] {10.1051/0004-6361/201117751},
  \href {https://ui.adsabs.harvard.edu/abs/2012A&A...537A.146E} {537, A146}

\bibitem[\protect\citeauthoryear{{Fossati} et~al.,}{{Fossati}
  et~al.}{2016}]{2016A&A...592A..84F}
{Fossati} L.,  et~al., 2016, \mn@doi [\aap] {10.1051/0004-6361/201628259},
  \href {https://ui.adsabs.harvard.edu/abs/2016A&A...592A..84F} {592, A84}

\bibitem[\protect\citeauthoryear{{Gaia Collaboration} et~al.,}{{Gaia
  Collaboration} et~al.}{2018}]{2018A&A...616A...1G}
{Gaia Collaboration} et~al., 2018, \mn@doi [\aap]
  {10.1051/0004-6361/201833051}, \href
  {https://ui.adsabs.harvard.edu/abs/2018A&A...616A...1G} {616, A1}

\bibitem[\protect\citeauthoryear{{Gonz{\'a}lez} \& {Levato}}{{Gonz{\'a}lez} \&
  {Levato}}{2006}]{2006A&A...448..283G}
{Gonz{\'a}lez} J.~F.,  {Levato} H.,  2006, \mn@doi [\aap]
  {10.1051/0004-6361:20053177}, \href
  {https://ui.adsabs.harvard.edu/abs/2006A&A...448..283G} {448, 283}

\bibitem[\protect\citeauthoryear{{Grunhut} et~al.,}{{Grunhut}
  et~al.}{2017}]{2017MNRAS.465.2432G}
{Grunhut} J.~H.,  et~al., 2017, \mn@doi [\mnras] {10.1093/mnras/stw2743}, \href
  {https://ui.adsabs.harvard.edu/abs/2017MNRAS.465.2432G} {465, 2432}

\bibitem[\protect\citeauthoryear{{Hempel} \& {Holweger}}{{Hempel} \&
  {Holweger}}{2003}]{2003AnA...408.1065H}
{Hempel} M.,  {Holweger} H.,  2003, \mn@doi [\aap]
  {10.1051/0004-6361:20030889}, \href
  {https://ui.adsabs.harvard.edu/abs/2003A&A...408.1065H} {408, 1065}

\bibitem[\protect\citeauthoryear{{Houk} \& {Swift}}{{Houk} \&
  {Swift}}{1999}]{1999MSS...C05....0H}
{Houk} N.,  {Swift} C.,  1999, Michigan Spectral Survey, \href
  {https://ui.adsabs.harvard.edu/abs/1999MSS...C05....0H} {5, 0}

\bibitem[\protect\citeauthoryear{{Jenkins} et~al.,}{{Jenkins}
  et~al.}{2016}]{2016SPIE.9913E..3EJ}
{Jenkins} J.~M.,  et~al., 2016, {The TESS science processing operations
  center}.
p. 99133E, \mn@doi{10.1117/12.2233418}

\bibitem[\protect\citeauthoryear{{Keszthelyi}, {Meynet}, {Georgy}, {Wade},
  {Petit}  \& {David-Uraz}}{{Keszthelyi} et~al.}{2019}]{2019MNRAS.485.5843K}
{Keszthelyi} Z.,  {Meynet} G.,  {Georgy} C.,  {Wade} G.~A.,  {Petit} V.,
  {David-Uraz} A.,  2019, \mn@doi [\mnras] {10.1093/mnras/stz772}, \href
  {https://ui.adsabs.harvard.edu/abs/2019MNRAS.485.5843K} {485, 5843}

\bibitem[\protect\citeauthoryear{{Kochukhov}, {Makaganiuk}  \&
  {Piskunov}}{{Kochukhov} et~al.}{2010}]{2010A&A...524A...5K}
{Kochukhov} O.,  {Makaganiuk} V.,   {Piskunov} N.,  2010, \mn@doi [\aap]
  {10.1051/0004-6361/201015429}, \href
  {https://ui.adsabs.harvard.edu/abs/2010A&A...524A...5K} {524, A5}

\bibitem[\protect\citeauthoryear{{Kochukhov} et~al.,}{{Kochukhov}
  et~al.}{2013}]{2013A&A...554A..61K}
{Kochukhov} O.,  et~al., 2013, \mn@doi [\aap] {10.1051/0004-6361/201321467},
  \href {https://ui.adsabs.harvard.edu/abs/2013A&A...554A..61K} {554, A61}

\bibitem[\protect\citeauthoryear{{Kupka}, {Piskunov}, {Ryabchikova}, {Stempels}
   \& {Weiss}}{{Kupka} et~al.}{1999}]{1999A&AS..138..119K}
{Kupka} F.,  {Piskunov} N.,  {Ryabchikova} T.~A.,  {Stempels} H.~C.,   {Weiss}
  W.~W.,  1999, \mn@doi [\aaps] {10.1051/aas:1999267}, \href
  {https://ui.adsabs.harvard.edu/abs/1999A&AS..138..119K} {138, 119}

\bibitem[\protect\citeauthoryear{{Kupka}, {Ryabchikova}, {Piskunov}, {Stempels}
   \& {Weiss}}{{Kupka} et~al.}{2000}]{2000BaltA...9..590K}
{Kupka} F.~G.,  {Ryabchikova} T.~A.,  {Piskunov} N.~E.,  {Stempels} H.~C.,
  {Weiss} W.~W.,  2000, \mn@doi [Baltic Astronomy] {10.1515/astro-2000-0420},
  \href {https://ui.adsabs.harvard.edu/abs/2000BaltA...9..590K} {9, 590}

\bibitem[\protect\citeauthoryear{{Landi Degl'Innocenti} \& {Landolfi}}{{Landi
  Degl'Innocenti} \& {Landolfi}}{2004}]{2004ASSL..307.....L}
{Landi Degl'Innocenti} E.,  {Landolfi} M.,  2004, {Polarization in Spectral
  Lines}.
~ Vol. 307, \mn@doi{10.1007/978-1-4020-2415-3, }

\bibitem[\protect\citeauthoryear{{Landstreet}, {Bagnulo}, {Andretta},
  {Fossati}, {Mason}, {Silaj}  \& {Wade}}{{Landstreet}
  et~al.}{2007}]{2007A&A...470..685L}
{Landstreet} J.~D.,  {Bagnulo} S.,  {Andretta} V.,  {Fossati} L.,  {Mason} E.,
  {Silaj} J.,   {Wade} G.~A.,  2007, \mn@doi [\aap]
  {10.1051/0004-6361:20077343}, \href
  {https://ui.adsabs.harvard.edu/abs/2007A&A...470..685L} {470, 685}

\bibitem[\protect\citeauthoryear{{Landstreet} et~al.,}{{Landstreet}
  et~al.}{2008}]{2008A&A...481..465L}
{Landstreet} J.~D.,  et~al., 2008, \mn@doi [\aap] {10.1051/0004-6361:20078884},
  \href {https://ui.adsabs.harvard.edu/abs/2008A&A...481..465L} {481, 465}

\bibitem[\protect\citeauthoryear{{Lomb}}{{Lomb}}{1976}]{1976Ap&SS..39..447L}
{Lomb} N.~R.,  1976, \mn@doi [\apss] {10.1007/BF00648343}, \href
  {http://adsabs.harvard.edu/abs/1976Ap%26SS..39..447L} {39, 447}

\bibitem[\protect\citeauthoryear{{Maitzen}}{{Maitzen}}{1976}]{1976A&A....51..223M}
{Maitzen} H.~M.,  1976, \aap, \href
  {https://ui.adsabs.harvard.edu/abs/1976A&A....51..223M} {51, 223}

\bibitem[\protect\citeauthoryear{{Maitzen} \& {Vogt}}{{Maitzen} \&
  {Vogt}}{1983}]{1983A&A...123...48M}
{Maitzen} H.~M.,  {Vogt} N.,  1983, \aap, \href
  {https://ui.adsabs.harvard.edu/abs/1983A&A...123...48M} {123, 48}

\bibitem[\protect\citeauthoryear{{Martioli}, {Teeple}  \& {Manset}}{{Martioli}
  et~al.}{2011}]{2011tfa..confE..63M}
{Martioli} E.,  {Teeple} D.,   {Manset} N.,  2011, in {Gajadhar} S.,  et~al.,
  eds, Telescopes from Afar. p.~63

\bibitem[\protect\citeauthoryear{{McDonald}, {Zijlstra}  \& {Boyer}}{{McDonald}
  et~al.}{2012}]{2012MNRAS.427..343M}
{McDonald} I.,  {Zijlstra} A.~A.,   {Boyer} M.~L.,  2012, \mn@doi [\mnras]
  {10.1111/j.1365-2966.2012.21873.x}, \href
  {https://ui.adsabs.harvard.edu/abs/2012MNRAS.427..343M} {427, 343}

\bibitem[\protect\citeauthoryear{{Monier}, {Griffin}, {Gebran},
  {K{\i}l{\i}{\c{c}}o{\u{g}}lu}, {Merle}  \& {Royer}}{{Monier}
  et~al.}{2019}]{2019AJ....158..157M}
{Monier} R.,  {Griffin} E.,  {Gebran} M.,  {K{\i}l{\i}{\c{c}}o{\u{g}}lu} T.,
  {Merle} T.,   {Royer} F.,  2019, \mn@doi [\aj] {10.3847/1538-3881/ab3b59},
  \href {https://ui.adsabs.harvard.edu/abs/2019AJ....158..157M} {158, 157}

\bibitem[\protect\citeauthoryear{{Munoz}, {Wade}, {Naz{\'e}}, {Puls}, {Bagnulo}
   \& {Szyma{\'n}ski}}{{Munoz} et~al.}{2020}]{2020MNRAS.492.1199M}
{Munoz} M.~S.,  {Wade} G.~A.,  {Naz{\'e}} Y.,  {Puls} J.,  {Bagnulo} S.,
  {Szyma{\'n}ski} M.~K.,  2020, \mn@doi [\mnras] {10.1093/mnras/stz2904}, \href
  {https://ui.adsabs.harvard.edu/abs/2020MNRAS.492.1199M} {492, 1199}

\bibitem[\protect\citeauthoryear{{Netopil}, {Paunzen}, {Maitzen}, {North}  \&
  {Hubrig}}{{Netopil} et~al.}{2008}]{2008A&A...491..545N}
{Netopil} M.,  {Paunzen} E.,  {Maitzen} H.~M.,  {North} P.,   {Hubrig} S.,
  2008, \mn@doi [\aap] {10.1051/0004-6361:200810325}, \href
  {https://ui.adsabs.harvard.edu/abs/2008A&A...491..545N} {491, 545}

\bibitem[\protect\citeauthoryear{{Netopil}, {Paunzen}, {H{\"u}mmerich}  \&
  {Bernhard}}{{Netopil} et~al.}{2017}]{2017MNRAS.468.2745N}
{Netopil} M.,  {Paunzen} E.,  {H{\"u}mmerich} S.,   {Bernhard} K.,  2017,
  \mn@doi [\mnras] {10.1093/mnras/stx674}, \href
  {https://ui.adsabs.harvard.edu/abs/2017MNRAS.468.2745N} {468, 2745}

\bibitem[\protect\citeauthoryear{{Pecaut} \& {Mamajek}}{{Pecaut} \&
  {Mamajek}}{2013}]{2013ApJS..208....9P}
{Pecaut} M.~J.,  {Mamajek} E.~E.,  2013, \mn@doi [\apjs]
  {10.1088/0067-0049/208/1/9}, \href
  {https://ui.adsabs.harvard.edu/abs/2013ApJS..208....9P} {208, 9}

\bibitem[\protect\citeauthoryear{{Petit} \& {David-Uraz}}{{Petit} \&
  {David-Uraz}}{2020}]{2020pase.conf..170P}
{Petit} V.,  {David-Uraz} A.,  2020, in {Wade} G.,  {Alecian} E.,  {Bohlender}
  D.,   {Sigut} A.,  eds, ~ Vol. 11, Stellar Magnetism: A Workshop in Honour of
  the Career and Contributions of John D. Landstreet. pp 170--177 (\mn@eprint
  {arXiv} {2004.04241})

\bibitem[\protect\citeauthoryear{{Petit} \& {Wade}}{{Petit} \&
  {Wade}}{2012}]{2012MNRAS.420..773P}
{Petit} V.,  {Wade} G.~A.,  2012, \mn@doi [\mnras]
  {10.1111/j.1365-2966.2011.20091.x}, \href
  {https://ui.adsabs.harvard.edu/abs/2012MNRAS.420..773P} {420, 773}

\bibitem[\protect\citeauthoryear{{Petit} et~al.,}{{Petit}
  et~al.}{2013}]{2013MNRAS.429..398P}
{Petit} V.,  et~al., 2013, \mn@doi [\mnras] {10.1093/mnras/sts344}, \href
  {https://ui.adsabs.harvard.edu/abs/2013MNRAS.429..398P} {429, 398}

\bibitem[\protect\citeauthoryear{{Piskunov}, {Kupka}, {Ryabchikova}, {Weiss}
  \& {Jeffery}}{{Piskunov} et~al.}{1995}]{1995A&AS..112..525P}
{Piskunov} N.~E.,  {Kupka} F.,  {Ryabchikova} T.~A.,  {Weiss} W.~W.,
  {Jeffery} C.~S.,  1995, \aaps, \href
  {https://ui.adsabs.harvard.edu/abs/1995A&AS..112..525P} {112, 525}

\bibitem[\protect\citeauthoryear{{Prv{\'a}k}, {Krti{\v{c}}ka}  \&
  {Korhonen}}{{Prv{\'a}k} et~al.}{2020}]{2020MNRAS.492.1834P}
{Prv{\'a}k} M.,  {Krti{\v{c}}ka} J.,   {Korhonen} H.,  2020, \mn@doi [\mnras]
  {10.1093/mnras/stz3564}, \href
  {https://ui.adsabs.harvard.edu/abs/2020MNRAS.492.1834P} {492, 1834}

\bibitem[\protect\citeauthoryear{{Renson} \& {Manfroid}}{{Renson} \&
  {Manfroid}}{2009}]{2009A&A...498..961R}
{Renson} P.,  {Manfroid} J.,  2009, \mn@doi [\aap]
  {10.1051/0004-6361/200810788}, \href
  {https://ui.adsabs.harvard.edu/abs/2009A&A...498..961R} {498, 961}

\bibitem[\protect\citeauthoryear{{Ricker} et~al.,}{{Ricker}
  et~al.}{2015}]{2015JATIS...1a4003R}
{Ricker} G.~R.,  et~al., 2015, \mn@doi [Journal of Astronomical Telescopes,
  Instruments, and Systems] {10.1117/1.JATIS.1.1.014003}, \href
  {https://ui.adsabs.harvard.edu/abs/2015JATIS...1a4003R} {1, 014003}

\bibitem[\protect\citeauthoryear{{Ryabchikova} \& {Romanovskaya}}{{Ryabchikova}
  \& {Romanovskaya}}{2017}]{2017AstL...43..252R}
{Ryabchikova} T.~A.,  {Romanovskaya} A.~M.,  2017, \mn@doi [Astronomy Letters]
  {10.1134/S1063773717040065}, \href
  {https://ui.adsabs.harvard.edu/abs/2017AstL...43..252R} {43, 252}

\bibitem[\protect\citeauthoryear{{Ryabchikova}, {Piskunov}, {Kupka}  \&
  {Weiss}}{{Ryabchikova} et~al.}{1997}]{1997BaltA...6..244R}
{Ryabchikova} T.~A.,  {Piskunov} N.~E.,  {Kupka} F.,   {Weiss} W.~W.,  1997,
  \mn@doi [Baltic Astronomy] {10.1515/astro-1997-0216}, \href
  {https://ui.adsabs.harvard.edu/abs/1997BaltA...6..244R} {6, 244}

\bibitem[\protect\citeauthoryear{{Ryabchikova}, {Piskunov}, {Kurucz},
  {Stempels}, {Heiter}, {Pakhomov}  \& {Barklem}}{{Ryabchikova}
  et~al.}{2015}]{2015PhyS...90e4005R}
{Ryabchikova} T.,  {Piskunov} N.,  {Kurucz} R.~L.,  {Stempels} H.~C.,  {Heiter}
  U.,  {Pakhomov} Y.,   {Barklem} P.~S.,  2015, \mn@doi [\physscr]
  {10.1088/0031-8949/90/5/054005}, \href
  {https://ui.adsabs.harvard.edu/abs/2015PhyS...90e4005R} {90, 054005}

\bibitem[\protect\citeauthoryear{{Scargle}}{{Scargle}}{1982}]{1982ApJ...263..835S}
{Scargle} J.~D.,  1982, \mn@doi [\apj] {10.1086/160554}, \href
  {http://adsabs.harvard.edu/abs/1982ApJ...263..835S} {263, 835}

\bibitem[\protect\citeauthoryear{{Schneider}, {Podsiadlowski}, {Langer},
  {Castro}  \& {Fossati}}{{Schneider} et~al.}{2016}]{2016MNRAS.457.2355S}
{Schneider} F.~R.~N.,  {Podsiadlowski} P.,  {Langer} N.,  {Castro} N.,
  {Fossati} L.,  2016, \mn@doi [\mnras] {10.1093/mnras/stw148}, \href
  {https://ui.adsabs.harvard.edu/abs/2016MNRAS.457.2355S} {457, 2355}

\bibitem[\protect\citeauthoryear{{Sch{\"o}ller} et~al.,}{{Sch{\"o}ller}
  et~al.}{2017}]{2017A&A...599A..66S}
{Sch{\"o}ller} M.,  et~al., 2017, \mn@doi [\aap] {10.1051/0004-6361/201628905},
  \href {https://ui.adsabs.harvard.edu/abs/2017A&A...599A..66S} {599, A66}

\bibitem[\protect\citeauthoryear{{Semel}}{{Semel}}{1989}]{1989A&A...225..456S}
{Semel} M.,  1989, \aap, \href
  {https://ui.adsabs.harvard.edu/abs/1989A&A...225..456S} {225, 456}

\bibitem[\protect\citeauthoryear{{Shultz} et~al.,}{{Shultz}
  et~al.}{2015}]{2015MNRAS.449.3945S}
{Shultz} M.,  et~al., 2015, \mn@doi [\mnras] {10.1093/mnras/stv564}, \href
  {https://ui.adsabs.harvard.edu/abs/2015MNRAS.449.3945S} {449, 3945}

\bibitem[\protect\citeauthoryear{{Shultz}, {Rivinius}, {Wade}, {Alecian},
  {Petit}  \& {BinaMIcS Collaboration}}{{Shultz}
  et~al.}{2018a}]{2018MNRAS.475..839S}
{Shultz} M.,  {Rivinius} T.,  {Wade} G.~A.,  {Alecian} E.,  {Petit} V.,
  {BinaMIcS Collaboration} 2018a, \mn@doi [\mnras] {10.1093/mnras/stx3238},
  \href {https://ui.adsabs.harvard.edu/abs/2018MNRAS.475..839S} {475, 839}

\bibitem[\protect\citeauthoryear{{Shultz} et~al.,}{{Shultz}
  et~al.}{2018b}]{2018MNRAS.475.5144S}
{Shultz} M.~E.,  et~al., 2018b, \mn@doi [\mnras] {10.1093/mnras/sty103}, \href
  {https://ui.adsabs.harvard.edu/abs/2018MNRAS.475.5144S} {475, 5144}

\bibitem[\protect\citeauthoryear{{Shultz} et~al.,}{{Shultz}
  et~al.}{2019a}]{2019MNRAS.485.1508S}
{Shultz} M.~E.,  et~al., 2019a, \mn@doi [\mnras] {10.1093/mnras/stz416}, \href
  {https://ui.adsabs.harvard.edu/abs/2019MNRAS.485.1508S} {485, 1508}

\bibitem[\protect\citeauthoryear{{Shultz} et~al.,}{{Shultz}
  et~al.}{2019b}]{2019MNRAS.490..274S}
{Shultz} M.~E.,  et~al., 2019b, \mn@doi [\mnras] {10.1093/mnras/stz2551}, \href
  {https://ui.adsabs.harvard.edu/abs/2019MNRAS.490..274S} {490, 274}

\bibitem[\protect\citeauthoryear{{Shultz} et~al.,}{{Shultz}
  et~al.}{2019c}]{2019MNRAS.490.4154S}
{Shultz} M.~E.,  et~al., 2019c, \mn@doi [\mnras] {10.1093/mnras/stz2846}, \href
  {https://ui.adsabs.harvard.edu/abs/2019MNRAS.490.4154S} {490, 4154}

\bibitem[\protect\citeauthoryear{{Sikora}, {Wade}, {Power}  \&
  {Neiner}}{{Sikora} et~al.}{2019a}]{2019MNRAS.483.3127S}
{Sikora} J.,  {Wade} G.~A.,  {Power} J.,   {Neiner} C.,  2019a, \mn@doi
  [\mnras] {10.1093/mnras/sty2895}, \href
  {https://ui.adsabs.harvard.edu/abs/2019MNRAS.483.3127S} {483, 3127}

\bibitem[\protect\citeauthoryear{{Sikora} et~al.,}{{Sikora}
  et~al.}{2019b}]{2019MNRAS.487.4695S}
{Sikora} J.,  et~al., 2019b, \mn@doi [\mnras] {10.1093/mnras/stz1581}, \href
  {https://ui.adsabs.harvard.edu/abs/2019MNRAS.487.4695S} {487, 4695}

\bibitem[\protect\citeauthoryear{{Soubiran}, {Le Campion}, {Brouillet}  \&
  {Chemin}}{{Soubiran} et~al.}{2016}]{2016AnA...591A.118S}
{Soubiran} C.,  {Le Campion} J.-F.,  {Brouillet} N.,   {Chemin} L.,  2016,
  \mn@doi [\aap] {10.1051/0004-6361/201628497}, \href
  {https://ui.adsabs.harvard.edu/abs/2016A&A...591A.118S} {591, A118}

\bibitem[\protect\citeauthoryear{{Stassun} et~al.,}{{Stassun}
  et~al.}{2018}]{2018AJ....156..102S}
{Stassun} K.~G.,  et~al., 2018, \mn@doi [\aj] {10.3847/1538-3881/aad050}, \href
  {https://ui.adsabs.harvard.edu/abs/2018AJ....156..102S} {156, 102}

\bibitem[\protect\citeauthoryear{{Stassun} et~al.,}{{Stassun}
  et~al.}{2019}]{2019AJ....158..138S}
{Stassun} K.~G.,  et~al., 2019, \mn@doi [\aj] {10.3847/1538-3881/ab3467}, \href
  {https://ui.adsabs.harvard.edu/abs/2019AJ....158..138S} {158, 138}

\bibitem[\protect\citeauthoryear{{Stibbs}}{{Stibbs}}{1950}]{1950MNRAS.110..395S}
{Stibbs} D.~W.~N.,  1950, \mn@doi [\mnras] {10.1093/mnras/110.4.395}, \href
  {https://ui.adsabs.harvard.edu/abs/1950MNRAS.110..395S} {110, 395}

\bibitem[\protect\citeauthoryear{{Thompson} et~al.,}{{Thompson}
  et~al.}{2012}]{2012ApJ...753...86T}
{Thompson} S.~E.,  et~al., 2012, \mn@doi [\apj] {10.1088/0004-637X/753/1/86},
  \href {https://ui.adsabs.harvard.edu/abs/2012ApJ...753...86T} {753, 86}

\bibitem[\protect\citeauthoryear{{Villebrun} et~al.,}{{Villebrun}
  et~al.}{2019}]{2019A&A...622A..72V}
{Villebrun} F.,  et~al., 2019, \mn@doi [\aap] {10.1051/0004-6361/201833545},
  \href {https://ui.adsabs.harvard.edu/abs/2019A&A...622A..72V} {622, A72}

\bibitem[\protect\citeauthoryear{{Vogt}, {Kerschbaum}, {Maitzen}  \&
  {Faundez-Abans}}{{Vogt} et~al.}{1998}]{1998A&AS..130..455V}
{Vogt} N.,  {Kerschbaum} F.,  {Maitzen} H.~M.,   {Faundez-Abans} M.,  1998,
  \mn@doi [\aaps] {10.1051/aas:1998238}, \href
  {https://ui.adsabs.harvard.edu/abs/1998A&AS..130..455V} {130, 455}

\bibitem[\protect\citeauthoryear{{Wade}, {Donati}, {Landstreet}  \&
  {Shorlin}}{{Wade} et~al.}{2000}]{2000MNRAS.313..851W}
{Wade} G.~A.,  {Donati} J.~F.,  {Landstreet} J.~D.,   {Shorlin} S.~L.~S.,
  2000, \mn@doi [\mnras] {10.1046/j.1365-8711.2000.03271.x}, \href
  {https://ui.adsabs.harvard.edu/abs/2000MNRAS.313..851W} {313, 851}

\bibitem[\protect\citeauthoryear{{Wade} et~al.,}{{Wade}
  et~al.}{2016a}]{2016MNRAS.456....2W}
{Wade} G.~A.,  et~al., 2016a, \mn@doi [\mnras] {10.1093/mnras/stv2568}, \href
  {https://ui.adsabs.harvard.edu/abs/2016MNRAS.456....2W} {456, 2}

\bibitem[\protect\citeauthoryear{{Wade}, {Petit}, {Grunhut}, {Neiner}  \&
  {MiMeS Collaboration}}{{Wade} et~al.}{2016b}]{2016ASPC..506..207W}
{Wade} G.~A.,  {Petit} V.,  {Grunhut} J.~H.,  {Neiner} C.,   {MiMeS
  Collaboration} 2016b, {Magnetic Fields of Be Stars: Preliminary Results from
  a Hybrid Analysis of the MiMeS Sample}.
p.~207

\bibitem[\protect\citeauthoryear{{Walborn}}{{Walborn}}{1972}]{1972AJ.....77..312W}
{Walborn} N.~R.,  1972, \mn@doi [\aj] {10.1086/111285}, \href
  {https://ui.adsabs.harvard.edu/abs/1972AJ.....77..312W} {77, 312}

\bibitem[\protect\citeauthoryear{{Walborn}}{{Walborn}}{1974}]{1974ApJ...191L..95W}
{Walborn} N.~R.,  1974, \mn@doi [\apjl] {10.1086/181558}, \href
  {https://ui.adsabs.harvard.edu/abs/1974ApJ...191L..95W} {191, L95}

\bibitem[\protect\citeauthoryear{{ud-Doula}, {Owocki}  \&
  {Townsend}}{{ud-Doula} et~al.}{2009}]{2009MNRAS.392.1022U}
{ud-Doula} A.,  {Owocki} S.~P.,   {Townsend} R. H.~D.,  2009, \mn@doi [\mnras]
  {10.1111/j.1365-2966.2008.14134.x}, \href
  {https://ui.adsabs.harvard.edu/abs/2009MNRAS.392.1022U} {392, 1022}

\bibitem[\protect\citeauthoryear{{ud-Doula}, {Owocki}  \& {Kee}}{{ud-Doula}
  et~al.}{2018}]{2018MNRAS.478.3049U}
{ud-Doula} A.,  {Owocki} S.~P.,   {Kee} N.~D.,  2018, \mn@doi [\mnras]
  {10.1093/mnras/sty1228}, \href
  {https://ui.adsabs.harvard.edu/abs/2018MNRAS.478.3049U} {478, 3049}

\makeatother
\end{thebibliography}




\appendix

\section{Binary fit for HD 25709}

In this section, we present the full binary solution that we obtain for HD~25709, assuming that the period of photometric variability is orbital in nature. We reemphasize however that this might not be the case, and that further observations are required to conclusively characterize this system.

\begin{table}
	\centering
	\caption{Orbital parameters of HD~25709: eccentricity ($e$), argument of periapsis ($\omega$), semi-amplitudes of the radial velocities ($K_1$ and $K_2$), systemic velocity ($\gamma$), epoch ($T_0$), mass ratio ($q \equiv M_2/M_1$), projected total mass, projected masses and projected semi-major axis ($a \sin i$).}
	\label{tab:appen}
	\begin{tabular}{l c} 
		\hline
		\multicolumn{2}{c}{Orbital parameters}\\
		\hline
		$e$ & $0.73\pm0.01$ \rule[-1.4ex]{0pt}{0pt}\\
		$\omega$ (\textdegree) & {$201\pm2$} \rule[-1.4ex]{0pt}{0pt}\\
		$K_1$ (km\ s$^{-1}$) &  {$72\pm2$} \rule[-1.4ex]{0pt}{0pt}\\
		$K_2$ (km\ s$^{-1}$) & {$-93^{+1}_{-2}$} \rule[-1.4ex]{0pt}{0pt}\\
		$\gamma$ (km\ s$^{-1}$) & {$18.4\pm0.5$} \rule[-1.4ex]{0pt}{0pt}\\
		$T_0$ & JD = {$2 458 807.585^{+0.008}_{-0.009}$} \rule[-1.4ex]{0pt}{0pt}\\
        {$q$} & {1.28$\pm$0.04} \rule[-1.4ex]{0pt}{0pt}\\
        $(M_1 + M_2) \sin^3 i$ ($\textrm{M}_\odot$) & $0.37\pm0.03$ \rule[-1.4ex]{0pt}{0pt}\\
        $M_1 \sin^3 i$ ($\textrm{M}_\odot$) & $0.21\pm0.01$ \rule[-1.4ex]{0pt}{0pt}\\
        $M_2 \sin^3 i$ ($\textrm{M}_\odot$) & $0.16\pm0.01$ \rule[-1.4ex]{0pt}{0pt}\\
        $a \sin i$ (AU) & $0.0263\pm0.0007$\\
		\hline
	\end{tabular}
\end{table}






\bsp	
\label{lastpage}
\end{document}